\documentclass[pra,aps,twoside,showpacs,superscriptaddress,twocolumn,floatfix,reprint,pdflatex]{revtex4-1}

\usepackage{color}

\usepackage{latexsym}
\usepackage{graphicx}
\usepackage{times}

\newcommand{\xxx}{}
\begin{document}

\title{Entangling efficiency of linear-optical quantum gates}

\author{Karel Lemr}
\author{Anton\'{\i}n \v{C}ernoch}
\author{Jan Soubusta}

\affiliation{Institute of Physics of Academy of Sciences of the Czech Republic, Joint Laboratory of Optics of PU and IP AS CR, 
   17. listopadu 50A, 779 07 Olomouc, Czech Republic}

\author{Miloslav Du\v{s}ek}

\affiliation{Department of Optics, Faculty of Science, Palack\'y University,
   17.~listopadu 12, 771 46 Olomouc, Czech~Republic}

\date{\today}

\begin{abstract}
We propose a new measure of non-classicality of quantum gates which is particularly suitable for probabilistic devices. This measure enables to compare, e.g., deterministic devices which prepare entangled states with low amount of entanglement with probabilistic devices which generate highly entangled states but which fail sometimes. We provide examples demonstrating advantages of this new measure over the so far employed entangling power.
\end{abstract}


\maketitle


\section{Introduction}

Quantum physics has opened new ways in information theory. Quantum computing and quantum information processing have attracted great attention in last few decades both in the theoretical and experimental domain \cite{nie00}. Quantum gates, the devices analogous to classical logical gates, are necessary ingredients for building quantum circuits. Considerable effort was devoted to implement them experimentally on several physical platforms that represent good candidates for this tasks \cite{gates-expNMR,gates-expCV,gates-expDV,gates-expATOM}. One of them is linear optics. The advantage of linear-optical quantum gates is that they are accessible by a present-day technology, their realization is relatively simple {\xxx\cite{simple}}, there is a chance of their integration \cite{integration}, and they work directly with light so they are convenient for information-processing tasks connected with quantum communication {\xxx\cite{gisin}}. The disadvantage is that they are mostly probabilistic, i.e. they operate with the success probability lower than one. This problem can be overcome by more complex setups based on pre-arranged entangled states \cite{kni01,cluster}. However, even the basic linear-optical quantum gates with success probability lower than one can be quite useful for small scale applications of quantum information processing. Especially for the circuits behind or between quantum links because in real-life quantum channels huge losses must be tolerated anyway.

It is an interesting question how to quantify the performance of quantum gates. It was proposed to use the so called \emph{entangling power} \cite{ent-pow1,ent-pow2,ent-pow3}. This measure is defined either as an average or maximal value of measures of entanglement of the output states over all separable input states. Of course, quantum gates are not primarily intended as entanglement sources (the sources of entangled states can be implemented more easily {\xxx\cite{kwiat}}) but their capability to create entanglement from separable input states is crucial for quantum information processing. Thus entangling power is a good measure of non-classicality of quantum gates.

However, this measure is disputable in the case of probabilistic quantum gates. One can consider two {\xxx distinct} cases: (i) a deterministic device which prepares entangled states with low amount of entanglement and (ii) a probabilistic device which generates highly entangled states but {\xxx which} fails sometimes. What is better? We can imagine we have a perfect entanglement-distillation apparatus which we apply to the output of the first device (i). We use it to obtain (asymptotically) such a number of distilled states which equals to the number of states generated by the second device (ii). The number of input states {\xxx is assumed to be} the same for both devices. Now, having the same fraction of states per a gate operation, we can find which of the two cases (i or ii) leads to a higher amount of entanglement. Thus a good measure of performance of probabilistic quantum gates could be the product of distillable entanglement \cite{dist-ent1,dist-ent2} and success probability maximized over all separable input states. But there are two problems with this definition. The first one is practical: It is difficult to calculate distillable entanglement for a general state. The second one is conceptual: This definition fails to quantify the ability to generate bound entangled states. Therefore it is convenient to generalize the definition of this quantity as the maximum (over all separable input states) of the product of success probability and any well-behaved entanglement measure chosen according to one's particular needs. We will call this function \emph{entangling efficiency}. It is reasonable to choose an entanglement measure which is convex because then the maximum can be taken only over all pure product input states.

In the following text we will define particular form of entangling efficiency using negativity as the measure of entanglement and we will use it to characterize the optimal linear-optical controlled phase gate recently implemented in our laboratory.

\section{Entangling efficiency}

In this paper we will measure the amount of entanglement in a quantum state by its negativity.
Negativity of state $\rho$ is defined as \cite{negativity}
\begin{equation}
\label{eq:def:neg}
N(\rho) = \frac{|| \rho^{T_A} ||_1 - 1}{2},
\end{equation}
where $|| \cdot ||_1$ denotes the trace norm and $T_A$ means the partial transpose.
This measure can be easily calculated. It is convex, $N(\sum_i p_i \rho_i) \leq \sum_i p_iN(\rho_i)$, and it is an entanglement monotone (does not change under local operation and classical communication). However, it is zero even if the state is entangled under positive partial transpose.

Entangling power is defined as the supremum of negativity of output states over all separable input states:
\begin{equation}
\label{eq:def:epow}
E_\mathrm{p}=  \sup_{\rho \in S} \left\{ N( \cal{E}[\rho] ) \right\},
\end{equation}
where $\cal{E}[\rho]$ denotes the output state of the device corresponding to input state $\rho$ ($\cal{E}$ is a completely positive map) and $S$ is the set of all separable input states.
Convexity of negativity is important, because for a non-convex entanglement measure there might be a mixed state whose measure of entanglement is greater than the negativity of any of the pure states it is {\xxx composed} of. This fact would complicate the search for the input state corresponding to the maximum because mixed states have significantly more degrees of freedom than pure states.

We define entangling efficiency in the following way:
\begin{equation}
\label{eq:def:eeff}
E_\mathrm{eff}= \sup_{\rho \in S} \left\{ p_\mathrm{s}(\rho) N( \cal{E}[\rho] ) \right\},
\end{equation}
where $p_\mathrm{s}(\rho)$ is the success probability of the gate for a given input state $\rho$.

\section{Example 1: Entangling efficiency of a beam splitter}

We illustrate the concept of entangling efficiency on an intuitive case of a beam splitter followed by a post-selection. Let us begin with a balanced beam splitter with transmittance $T$ and reflectance $R$ both equal to $1 \over 2$. Two photonic qubits are initially in separable states and each of them enters one input port of the beam splitter. Subsequently we perform a post-selection taking into account only the cases where there is exactly one photon in each output mode.

Let us express a separable input state, $|\psi_1\rangle \otimes |\psi_2\rangle$, using the following parametrization (without the loss of generality one part of the input state can be fixed):
\begin{eqnarray}
|\psi_{1}\rangle &=& |0\rangle \nonumber\\
|\psi_{2}\rangle &=& \cos{\theta_{}}|0\rangle + e^{i\vartheta_{}}\sin{\theta_{}}|1\rangle,
\end{eqnarray}
where $\lbrace|0\rangle,|1\rangle\rbrace$ represents an arbitrary orthogonal basis. Simple algebra reveals that the success probability of the above described beam-splitter transformation reads
\begin{equation}
p_\mathrm{s}(\theta) = \frac{\sin^2\theta}{2}.
\end{equation}
Using definition (\ref{eq:def:neg}) one can calculate negativity of output states
\begin{equation}
N = \frac{1}{2},
\end{equation}
which is independent of the input state. Therefore the entangling power of the beam splitter reads:
\begin{equation}
E_\mathrm{p} = \frac{1}{2}.
\end{equation}
As for the entangling efficiency, one has to find the maximum of the product of success probability and negativity
\begin{equation}
N p_\mathrm{s} = \frac{\sin^2\theta}{4}.
\end{equation}
This product is maximized for $\theta=\pi/2$ so the entangling efficiency is
\begin{equation}
E_\mathrm{eff} = \frac{1}{4}.
\end{equation}
\begin{figure}
\includegraphics[scale=1]{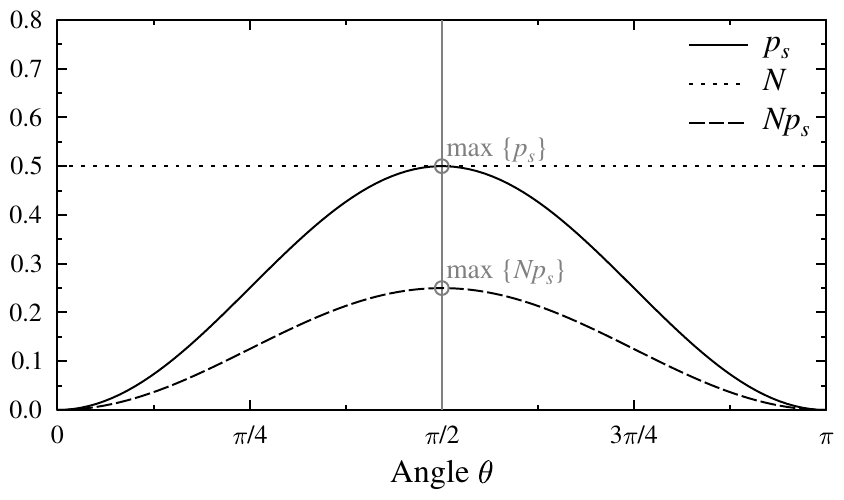}
\caption{Balanced beamsplitter. Success probability $p_\mathrm{s}$ {\xxx (full line)}, negativity $N$ {\xxx (dotted line)} and their product {\xxx (dashed line)} as functions of angle $\theta$ parametrizing the input separable state.}
\label{fig:bs_state}
\end{figure}
Fig.~\ref{fig:bs_state} shows the dependence of success probability $p_\mathrm{s}$, negativity $N$ and their mutual product as functions of angle $\theta$. Finding the maximum of negativity $N$ reveals the entangling power, $E_\mathrm{p}$, of the beam splitter, whereas finding the maximum of the product $N p_\mathrm{s}$ reveals the entangling efficiency, $E_\mathrm{eff}$. Note that in this case, both of them are maximized for the same input state parameter $\theta$.

\begin{figure}
\includegraphics[scale=1]{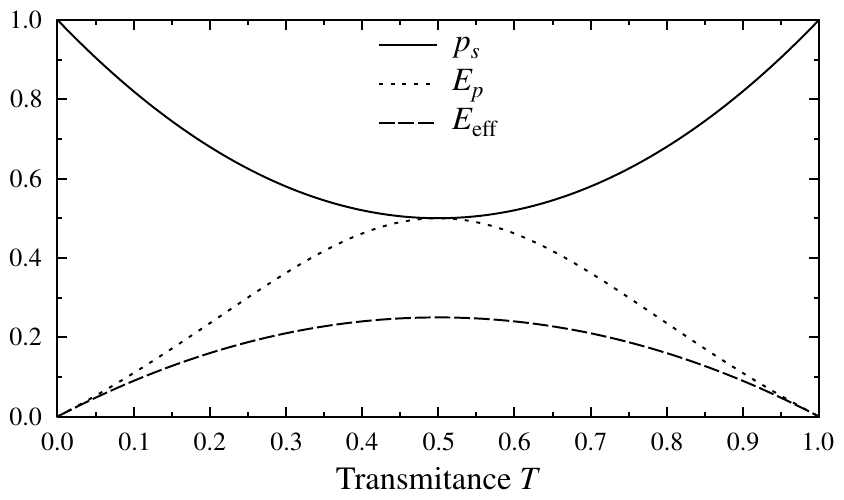}
\caption{Beamsplitter. Success probability $p_\mathrm{s}$ {\xxx (full line)}, entangling power $E_\mathrm{p}$ {\xxx (dotted line)} and entangling efficiency $E_\mathrm{eff}$ {\xxx (dashed line)} of a beam splitter as functions of its transmittance $T$.}
\label{fig:bs_t}
\end{figure}
One can extend the model of balanced beam splitter also to the case of a general lossless beam splitter. We have calculated the success probability $p_\mathrm{s}$, entangling power $E_\mathrm{p}$ and entangling efficiency $E_\mathrm{eff}$ also for this case. Fig.~\ref{fig:bs_t} presents {\xxx these} three properties of the beam splitter as functions of its transmittance.

\section{Example 2: Optimal linear-optical c-phase gate} \label{sec:cgate}

We further demonstrate our concept of entangling power on the second example: The optimal linear-optical controlled-phase (c-phase) gate. A controlled phase gate implements the following operation on two qubits:
\begin{equation}
\begin{array}{lcl}
 |0, 0 \rangle &\mapsto&  |0, 0 \rangle , \\
 |0, 1 \rangle &\mapsto&  |0, 1 \rangle , \\
 |1, 0 \rangle &\mapsto&  |1, 0 \rangle , \\
 |1, 1 \rangle &\mapsto&  e^{i \varphi}|1, 1 \rangle.
\end{array}
\label{eqnn:gate}
\end{equation}
In general, it is an entangling quantum gate. Together with single-qubit operations, it forms a universal set for quantum computing. E.g., the controlled-NOT gate can be obtained by applying a Hadamard transform to the target qubit before and after the controlled-phase gate with phase shift $\pi$.

\begin{figure}
  \begin{center}
  \resizebox{\hsize}{!}{\includegraphics*{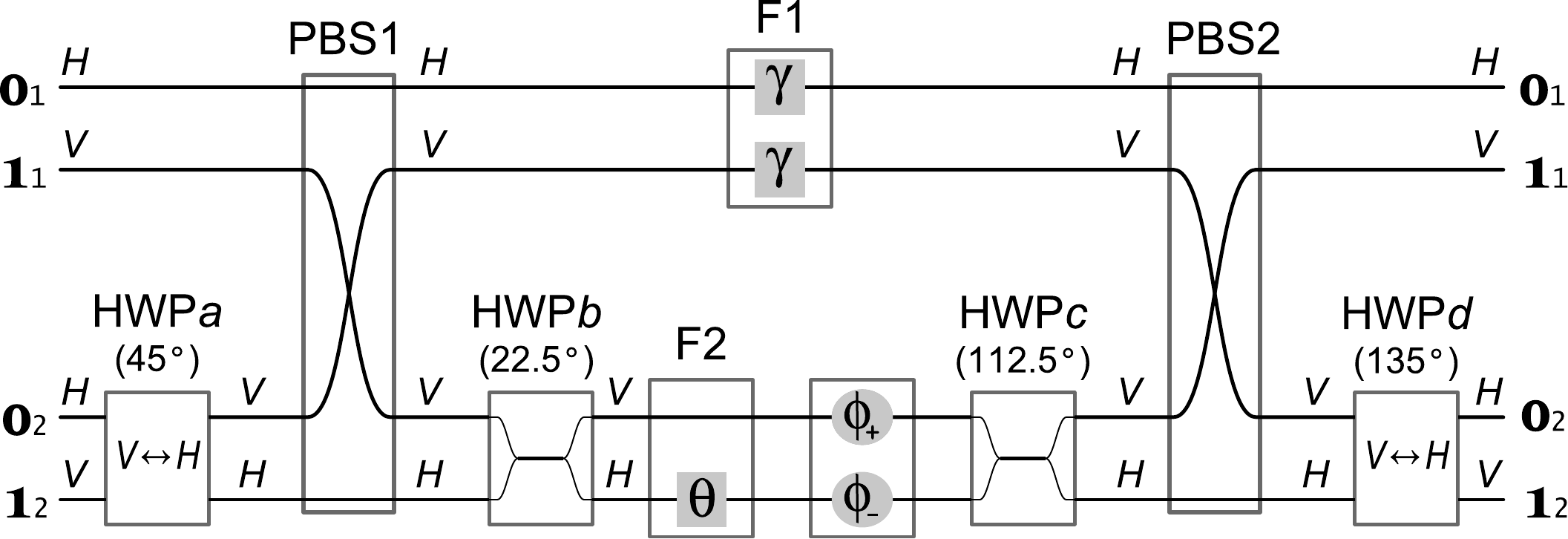}}
  \end{center}
  \caption{Scheme of the gate \cite{my}. Vertically ($V$)
  and horizontally ($H$) polarized components of the same beam
  are drawn separately for clarity. In polarization beam splitters
  PBS1 and PBS2 the vertical components are reflected. Half-wave
  plates HWP$b$ and HWP$c$ act as ``beam splitters'' for $V$ and $H$
  polarization modes. F1 and F2 are filters (attenuators),
  F1 acts on the both polarization modes,
  F2 on the $H$ component only.
  Phase shifts $\phi_{+}$ and $\phi_{-}$ are introduced by proper
  path differences in the respective modes.
  HWP$a$ and HWP$d$ just swap vertical and horizontal
  polarizations. In the final setup they are omitted for simplicity
  and the second qubit is encoded inversely with respect to the first qubit.}
  \label{scheme1}
\end{figure}

\begin{figure}
\includegraphics[scale=1]{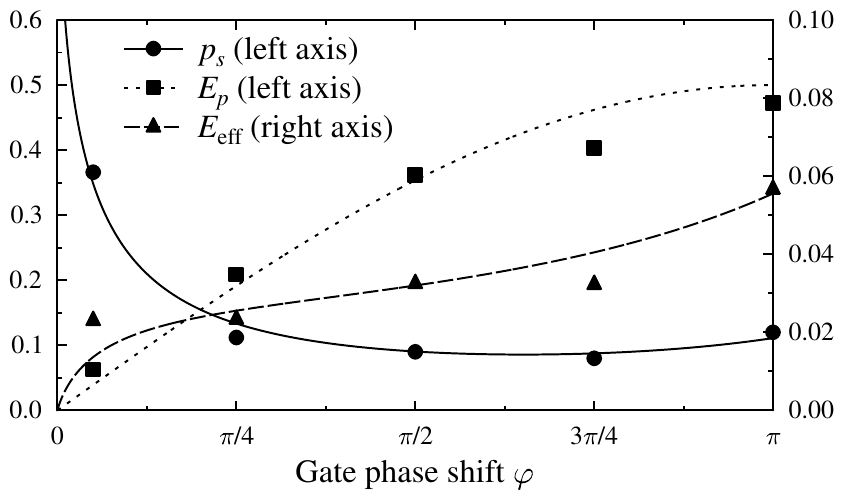}
\caption{C-phase gate. Success probability $p_\mathrm{s}$ {\xxx (full line)}, entangling power $E_\mathrm{p}$ {\xxx (dotted line)}, and entangling efficiency $E_\mathrm{eff}$ {\xxx (dashed line)} as functions of phase shift $\varphi$ applied by the gate. Theoretical prediction is depicted by lines whereas the experimentally obtained data are depicted using markers.}
\label{fig:gate}
\end{figure}

Recently, we built the optimal linear-optical controlled phase gate in our laboratory \cite{my}. Its conceptual scheme is {\xxx depicted} in Fig.~\ref{scheme1}. Phase shift $\varphi$ applied by this gate on the controlled qubit can be set to any given value just by tuning the parameters of the setup. The gate is optimal in the sense that for any phase shift it operates at the maximum possible success probability that is achievable within the framework of any post-selected linear-optical implementation without auxiliary photons. The optimal success probability of the gate takes the following form \cite{kie10}
\begin{eqnarray}
   p_\mathrm{s}(\varphi) =
   \left(1+2\left|\sin\frac{\varphi}{2}\right|+2^{3/2}\sin\frac{\pi-\varphi}{4}{\left|\sin\frac{\varphi}{2}\right|^{1/2}}\right)^{-2}
\label{eqnn:psucc}
\end{eqnarray}
The dependence of the success probability on the phase shift is shown on Fig.~\ref{fig:gate}. Surprisingly it is \emph{not monotone} in the phase.
\begin{figure}
\includegraphics[scale=1]{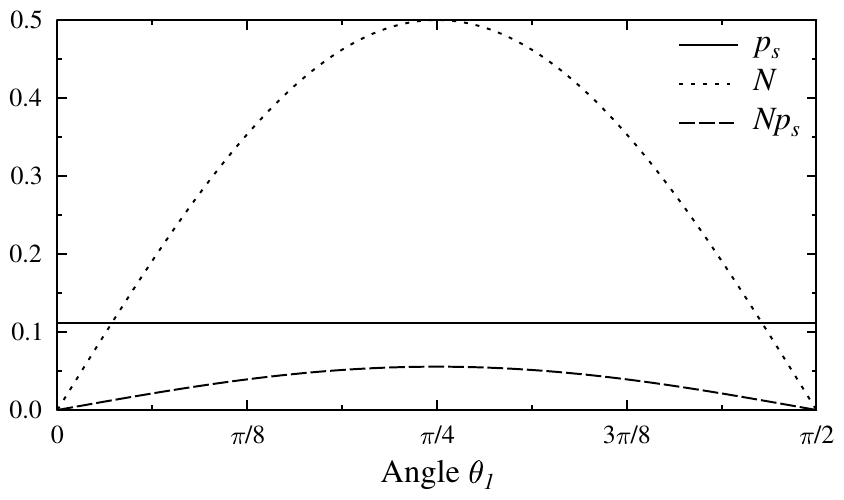}
\caption{C-phase gate. Success probability $p_\mathrm{s}$ (full line), negativity $N$ (dotted line) and their mutual product (dashed line) as functions of the parameter $\theta_1$ for fixed value of $\varphi = \pi$ and $\theta_2 = \frac{\pi}{4}$.}
\label{fig:gate_state}
\end{figure}

In order to evaluate the entangling power and efficiency of this gate, let us express separable input state $|\psi_1\rangle \otimes |\psi_2\rangle$ using the following parametrization
\begin{equation}
\label{eq:gate_inp_state}
|\psi_{1,2}\rangle = \cos{\theta_{1,2}}|0\rangle + e^{i\vartheta_{1,2}}\sin{\theta_{1,2}}|1\rangle,
\end{equation}
where $\lbrace|0\rangle,|1\rangle\rbrace$ represents a fixed computational basis {\xxx and indices denote the first and second qubit}. Further, let us assume this state is successfully transformed by the gate according to Eq.~(\ref{eqnn:gate}). Then we can calculate negativity of the output two-qubit state using Eq.~(\ref{eq:def:neg}):
\begin{equation}
\label{eq:neg_cphase}
N(\varphi,\theta_1,\theta_2) = \frac{\sin{2\theta_1}}{2}\frac{\sin{2\theta_2}}{2}\sqrt{2(1-\cos{\varphi})},
\end{equation}
where $\varphi$ denotes the phase applied by the gate. One can notice that this function does not depend on phases $\vartheta_{1,2}$ but only on $\theta_{1,2} \in [0, \frac{\pi}{2}]$. For any given value of gate phase shift $\varphi$, it is maximized for $\theta_{1,2} = \frac{\pi}{4}$ (equal superposition of computational basis states). The maximum negativity and therefore the entangling power for a given phase $\varphi$ thus reads
\begin{equation}
\label{eq:gate_ep}
E_\mathrm{p}(\varphi) = \frac{\sqrt{2}}{4}\sqrt{1-\cos{\varphi}}.
\end{equation}
Similarly, the entangling efficiency can be obtained by maximizing the product
$ N(\varphi,\theta_1,\theta_2)\, p_\mathrm{s}(\varphi) $
over the input-state parameters $\theta_{1,2}$. Because the success probability does not depend on the input state, we obtain the entangling efficiency as a function of the gate phase shift, $\varphi$, in the form
\begin{equation}
E_\mathrm{eff} = \frac{\sqrt{2}\, p_\mathrm{s}}{4}\sqrt{1-\cos{\varphi}}.
\end{equation}
Clearly, the success probability of our c-phase gate is state independent [see Eq.~(\ref{eqnn:psucc})], so entangling efficiency of the gate is just a product of success probability and entangling power.

In Fig.~{\ref{fig:gate_state}} there are plots of success probability $p_\mathrm{s}$, negativity $N$, and their product in dependence {\xxx on} parameter $\theta_1$ for $\varphi = \pi$ and $\theta_2=\frac{\pi}{4}$ (this parameter is kept fixed and equal to its optimal value). Fig.~{\ref{fig:gate}} shows success probability $p_\mathrm{s}$, entangling power $E_\mathrm{p}$, and entangling efficiency $E_\mathrm{eff}$ as functions of $\varphi$.

Because of the high importance of the c-phase gate for quantum computation, we have tested its entangling power and efficiency also experimentally. To calculate entangling power and efficiency from experimental data we scanned the four-parametric space of all separable input states numerically (over $40\,000$ uniformly distributed states were tested). The corresponding output states were calculated using the Choi matrices, $\chi(\varphi)$, reconstructed by means of the quantum process tomography \cite{my,Poyatos97}, using the following formula \cite{cho75}: $\rho_\mathrm{out} = \mbox{Tr}_\mathrm{in} \left[ \chi(\varphi) \left( \rho_\mathrm{in}^T\otimes 1 \right) \right]$.
Entangling power and other quantities are shown in Fig.~\ref{fig:gate} where the values obtained from experimental data can be compared with theoretical predictions.

\section{Example 3: Generalized c-phase gate}

From the two previous examples the reader may get impression that entangling efficiency does not provide any substantial benefit with respect to entangling power. The reason lies in the specific nature of these examples. In the case of the beam splitter, negativity is independent on the input state, so it can be factored out of the maximum search in the formula for entangling efficiency [see Eq.~(\ref{eq:def:eeff})]. In the case of the c-phase gate the probability of success does not depend on the input state, so it can be taken out of the maximization. Therefore in both these situations the entangling power gives qualitatively the same results as the entangling efficiency gives.

\begin{figure}
\includegraphics[scale=1]{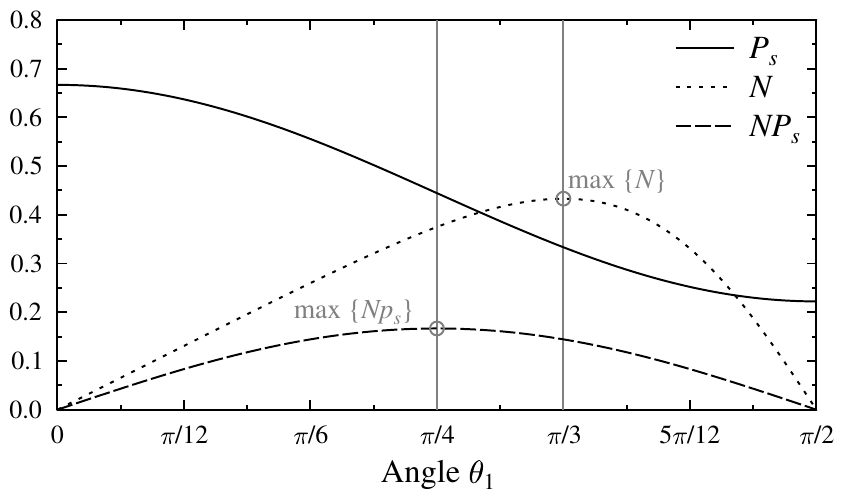}
\caption{Generalized c-phase gate. Success probability $P_\mathrm{s}$ {\xxx (full line)}, negativity $N$ {\xxx (dotted line)}, and their mutual product {\xxx (dashed line)} are plotted against the parameter of input state $\theta_1$ for $\varphi=\pi$ and $\theta_2 = \pi/4$. All the functions are symmetric in the sense that the same results would be obtained when exchanging $\theta_1$ and $\theta_2$.}
\label{fig:gcpg_state}
\end{figure}
\begin{figure}
\includegraphics[scale=1]{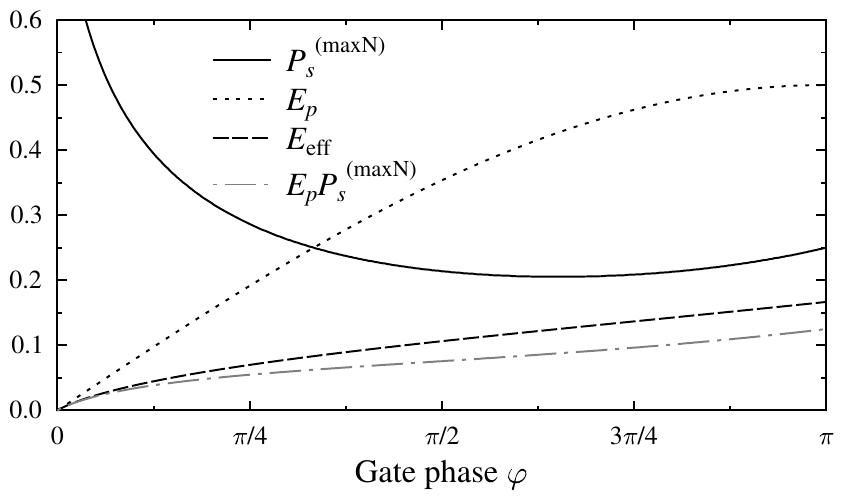}
\caption{Generalized c-phase gate. Entangling power $E_\mathrm{p}$ {\xxx (dotted line)}, corresponding success probability $P_\mathrm{s}^\mathrm{(maxN)}$ {\xxx (full line)}, their product {\xxx (dotted-dashed grey line)}, and entangling efficiency $E_\mathrm{eff}$ {\xxx (dashed line)} are plotted against gate phase shift $\varphi$.}
\label{fig:gcpg_phi}
\end{figure}

Here we expose the third example, the generalized c-phase gate, which proves that the entangling efficiency is, in general, better instrument than the entangling power. The linear-optical scheme from Fig.~\ref{scheme1} performs the c-phase gate transformation (\ref{eqnn:gate}) only if a compensating filter, F1, with a proper transmissivity ($\gamma = p_\mathrm{s}^{1/4}$) is used in the upper path. If this filter is removed the device will perform the following generalized (non-unitary) transformation:
\begin{equation}
\begin{array}{lcl}
 |0, 0 \rangle &\mapsto&  |0, 0 \rangle , \\
 |0, 1 \rangle &\mapsto&  \sqrt[4]{p_\mathrm{s} (\varphi)} |0, 1 \rangle , \\
 |1, 0 \rangle &\mapsto&  \sqrt[4]{p_\mathrm{s} (\varphi)} |1, 0 \rangle , \\
 |1, 1 \rangle &\mapsto&  \sqrt{p_\mathrm{s} (\varphi)} e^{i \varphi} |1, 1 \rangle.
\end{array}
\label{eqnn:gate2}
\end{equation}
We will call it a \emph{generalized c-phase gate}. Such gate can be used for instance in quantum nondemolition measurement {\xxx\cite{qnd}}. Quantity $p_\mathrm{s}(\varphi)$ is defined by Eq.~(\ref{eqnn:psucc}) but it does not represent the success probability of the generalized gate. The overall success probability, $P_\mathrm{s}$, is now a function of the input state
\begin{eqnarray}
\label{eq:gcpg-psucc}
P_\mathrm{s}(\varphi, \theta_1, \theta_2) & = & \cos^2 \theta_1 \cos^2 \theta_2 \nonumber\\
& + & \sqrt{p_\mathrm{s} (\varphi)}(\cos^2 \theta_1 \sin^2 \theta_2 + \sin^2 \theta_1 \cos^2 \theta_2)  \nonumber\\
& + & p_\mathrm{s}(\varphi) \sin^2 \theta_1 \sin^2 \theta_2.
\end{eqnarray}
Assuming again pure product states parametrized by Eq.~(\ref{eq:gate_inp_state}) in the input, simple calculation reveals the formula for negativity of corresponding output states:
\begin{equation}
\label{eq:gcpg-neg}
N(\varphi, \theta_1,\theta_2) = \frac{\sin 2\theta_1 \sin 2\theta_2}{2\sqrt{2}}\frac{
\sqrt{p_\mathrm{s}(\varphi)}\sqrt{1-\cos\varphi}}{P_\mathrm{s}(\varphi, \theta_1, \theta_2)} .
\end{equation}
It can be seen (look at Fig.~\ref{fig:gcpg_state}) that the negativity alone is now maximized for generally different angles $\theta_1$ and $\theta_2$ than in the case of the ``standard'' c-phase gate (Sec.~\ref{sec:cgate}). The product of negativity and success probability, however, finds its maximum still in $\pi/4$. Fig.~\ref{fig:gcpg_state} presents success probability, negativity and their product as functions of $\theta_1$ for $\varphi = \pi$ and $\theta_2 = \pi/4$. Note that the relation is symmetric for $\theta_2$. In this figure, one can clearly perceive that now negativity is maximized for different parameters than the product of negativity and success probability.

Fig.~\ref{fig:gcpg_phi} is similar to Fig.~\ref{fig:gate} showing how entangling power and entangling efficiency varies with the phase shift $\varphi$. Because in the case of this gate the success probability is state dependent we plot here success probability $P_\mathrm{s}^\mathrm{(maxN)}$ corresponding to the states which maximize the negativity. Besides, we have added function $P_\mathrm{s}^\mathrm{(maxN)} E_\mathrm{p}$. This function should help to view two measures of non-classicality of the gate, entangling power and entangling efficiency, under comparable conditions. {\xxx One can perceive that entangling efficiency is greater then the product $P_\mathrm{s}^\mathrm{(maxN)} E_\mathrm{p}$, because it takes the success probability into the maximization.}

\section{Conclusions}

There is no doubt that a concept of a measure of non-classicality of quantum gates, which takes into account the success probability, is more natural for probabilistic devices than, e.g., the concept of entangling power. The question was if such a measure, namely the entangling efficiency defined above in this paper, can really offer different and more appropriate information in the case of linear optical devices than the entangling power. Our last example shows that it can. In general, entangling efficiency is not a trivial function of entangling power.
It indicates that entangling efficiency is a useful measure of entangling capability of probabilistic quantum gates and that entangling power may sometimes yield deficient information about this capability.

\begin{acknowledgments}
This work was supported by Palack\'{y} University (PrF-2012-019), by Institute of Physics of the Czech Academy of Sciences (AVOZ10100522) and by the Czech Science Foundation (P205/12/0382).
\end{acknowledgments}

\end{document}